\documentstyle[12pt]{article}
\begin{document}
\newcommand{\eqn}[1]{(\ref{eq:#1})}
\newcommand{\ben}{\begin{equation}}
\newcommand{\een}{\end{equation}}
\newcommand{\bea}{\begin{eqnarray}}
\newcommand{\eea}{\end{eqnarray}}
\newcommand{\nn}{\nonumber \\ }
\newcommand{\bdm}{\begin{displaymath}}
\newcommand{\edm}{\end{displaymath}}
\newcommand{\A}{\cal{A}_\gamma }
\newcommand{\At}{\tilde{\cal{A}}_{\gamma}}
\newcommand{\Ak}{{\cal A}_{3k}}
\newcommand{\SU}{\widehat{SU(2)}}
\newcommand{\SP}{\widehat{SU(2)^+}}
\newcommand{\SM}{\widehat{SU(2)^-}}
\newcommand{\SSS}{\widehat{SU(3)}_{\tilde{k}^+}}
\newcommand{\br}{\langle}
\newcommand{\kt}{\rangle}
\newcommand{\bra}[1]{\langle {#1}|}
\newcommand{\ket}[1]{|{#1}\rangle}
\newcommand{\lm}{\ell^-}
\newcommand{\lp}{\ell^+}
\newcommand{\la}{\lambda}
\newcommand{\al}{\alpha}
\newcommand{\eps}{\epsilon}
\newcommand{\vl}{\vec{\lambda}}
\newcommand{\hf}{\frac{1}{2}}
\newcommand{\pa}{\partial}
\newcommand{\ktp}{\tilde{k}^+}
\newcommand{\ktm}{\tilde{k}^-}
\newcommand{\CMP}{Commun. Math. Phys. }

\begin{titlepage}
\null
\begin{flushright}
NBI-HE-92-43 \\
July 1992
\end{flushright}
\vspace{1cm}
\begin{center}
{\Large\bf
Free Field Representations
and Screening Operators \\
for the $N=4$ Doubly Extended
Superconformal Algebras
\par}
\vskip 3em
\lineskip .75em
\normalsize
\begin{tabular}[t]{c}
{\bf Katsushi Ito},
{\bf Jens Ole Madsen}  {\bf and}
{\bf Jens Lyng Petersen}  \\ \\
{\sl The Niels Bohr Institute}, {\sl Blegdamsvej 17} \\
{\sl DK-2100}, {\sl Copenhagen {\O}, Denmark}
\end{tabular}
\vskip 1.5em
{\bf Abstract}
\end{center} \par
We present explicit free field representations for the $N=4$ doubly
extended superconformal algebra, $\tilde{\cal{A}}_{\gamma}$.
This algebra generalizes and contains all previous $N=4$ superconformal
algebras. We have found $\tilde{\cal{A}}_{\gamma}$ to be obtained by
hamiltonian reduction of the Lie superalgebra  $D(2|1;\alpha)$.
In addition, screening operators are explicitly given and the associated
singular vectors identified. We use this to present a natural conjecture
for the Kac determinant generalizing a previous conjecture by Kent and
Riggs for the singly extended case. The results support and
illuminate several aspects of the characters of this algebra previously
obtained by Taormina and one of us.

\end{titlepage}
\baselineskip=0.8cm
\renewcommand{\thefootnote}{\arabic{footnote}}
\setcounter{footnote}{0}
In recent works, we \cite{imp} and two of us \cite{im}, have outlined
how essentially all known superconformal algebras -- including
(super-$W$-type) ones (for $N\ge4$) for which the algebra only closes
on composites of generators -- may be classified and indeed be given
a unified treatment, using the techniques of hamiltonian reductions of
WZNW models on Lie (super)groups \cite{hamred}.

In ref.\cite{im} this program was described at the level of the
{\em classical} hamiltonian reduction, whereas in ref. \cite{imp} the
quantum case was considered as well using explicit free field
realizations. In the case of superconformal algebras with an affine
$\widehat{SO}(N)$ algebra such a realization has been known for some
time \cite{mathieu,miki}. In \cite{imp} it was further discussed and
explicit realizations were described for the non-exceptional type Lie
superalgebras, $sl(N|2)$, $osp(N|2)$ and $osp(4|2N)$.
For the $N=4$ singly extended  algebra containing an affine
$\widehat{SU}(2)$ \cite{ademollo} it has been recently provided by
Matsuda, \cite{matsuda}.
In the present letter we describe the case of the $N=4$ ``doubly extended"
superconformal algebra of Sevrin, Troost and Van Proeyen \cite{stvp88}
(see also \cite{ivan}-\cite{s89}), commonly denoted $\At$
(see \cite{gptvp}-\cite{gs}) and further analyzed by several authors
(\cite{pt}-\cite{defev}).
The algebra, $\At$, was originally given in a form with auxiliary free
fermionic generators and an extra $U(1)$ current.
That algebra is denoted $\A$ and
has the superalgebra $D(2|1;\alpha)$ as its finite
dimensional sub-algebra. We have found that $\At$ is obtained as a
result of carrying out the hamiltonian reduction of the WZNW model on the
Lie supergroup of the Lie superalgebra $D(2|1;\alpha)$.

We first provide a brief account of how to translate between the notations
of \cite{imp} using the Lie superalgebra language, and the original
notations for $\At$ (see for example \cite{stvp88,gptvp}).

The exceptional Lie superalgebra, $D(2|1;\alpha)$ (cf. ref. \cite{Kac})
depends on the parameter $\alpha$ or $\gamma\equiv\frac{\alpha}{1-\alpha}$,
and has rank three. Define the simple roots, $\alpha_1,\alpha_2,\alpha_3$
with  metric given by
\bea
\alpha_1^2=0,&\alpha_2^2=-2\gamma, &\alpha_3^2=-2(1-\gamma),\nn
\alpha_1\cdot\alpha_2=\gamma,&\alpha_1\cdot\alpha_3=1-\gamma,&
\alpha_2\cdot\alpha_3=0,
\eea
Then the positive (mutually orthogonal) even roots are
$\alpha_2$, $\alpha_3$ and
$\theta\equiv 2\alpha_1+\alpha_2+\alpha_3\equiv\alpha_{1123}$,
with $\theta^2=2$.
They define the even subalgebra, ${\bf g}_{0}=A_1\oplus
A_1\oplus A_1$, or (partly in the notation
of \cite{stvp88}), $su(2)^+\oplus su(2)^-\oplus su(2)_\theta$.
$\alpha_2,\alpha_3$ define the set $\Delta^0_+(G)$ with
$G=su(2)^+\oplus su(2)^-$ in the notation of \cite{imp}.
It is the affine $\widehat{SU}(2)_\theta$ algebra defined by $\theta$
which is constrained in the hamiltonian reduction and gives rise to the
Virasoro subalgebra.

The positive odd roots forming the set $\Delta^1_+$ are given by
\bea
\alpha_1&\equiv&{\beta_-} ,\quad\quad
\alpha_{12}\equiv\alpha_1+\alpha_2\equiv\beta_{+K}, \nn
\alpha_{13}&\equiv&\alpha_1+\alpha_3\equiv\beta_{-K}, \quad
\alpha_{123}\equiv\alpha_1+\alpha_2+\alpha_3\equiv\beta_+,
\eea
the $+,-,+K,-K$ notation being used in the original literature on $\At$.
The pairing $+\leftrightarrow -$ and $+K\leftrightarrow -K$ corresponds
to the pairing
$\beta\leftrightarrow\theta-\beta$ (with $\beta,\theta-\beta\in\Delta_+^1$)
in the notation of \cite{imp}.
The vector space, ${\bf g}_{1}$, of the odd generators carry a
representation of $G=su(2)^+\oplus su(2)^-$ the representation matrices of
which may be obtained as follows:

For  $\beta,\beta'\in\Delta_+^1$, and
$\alpha\equiv \beta'-\beta\in \Delta^0$, define
\ben
t^{\alpha}_{\theta-\beta',\beta} \equiv
2\frac{N_{-\beta',\beta}}{\alpha^2}
=N_{\alpha ,\beta}
\een
where $N_{\beta',\beta}$ are the structure coefficients of (the odd-odd
part of) $D(2|1;\alpha)$.
The relation to the notation of \cite{stvp88}  consists in denoting
$$+\alpha_2 : (++), \ \ \ -\alpha_2 : (+-),\ \ \ +\alpha_3 : (-+),
\ \ \ -\alpha_3 : (--)$$
One then finds
\ben
t_{+,+K}^{+-}= +1,\quad t_{-,-K}^{++}=-1,\quad
t_{-,+K}^{-+}=-1, \quad t_{+,-K}^{--}=+1,
\een
the remaining ones being zero. Matrix multiplication is defined by
\ben
(t^{\alpha}t^{\alpha'})_{\theta-\beta',\beta}
\equiv \sum_{\beta''\in\Delta^1_+}t_{\theta-\beta',\beta''}^{\alpha}
t_{\theta-\beta'',\beta}^{\alpha'}.
\een
Define $t^2$ and $t^3$
(corresponding to $2t^{+3}$ and $2t^{-3}$ in the traditional notation) by
$ [t^{\alpha_i},t^{-\alpha_i}]=t^i. $
Then
\ben
 \pm t^{\pm 3}_{+K,-K}=\mp t^{\pm 3}_{-K,+K}= \hf ,\quad \quad
t^{\pm 3}_{+,-}=-t^{\pm 3}_{-,+}= \hf .
\een
Thus we have the two commuting doublet representations of the two $SU(2)$'s
($i=2,3$):
$$\vec{t}_i\equiv (t^{\alpha_i},t^{-\alpha_i},t^i)=
(t^{\pm +},t^{\pm -},2t^{\pm 3})\equiv \vec{t}^\pm.$$
The space, ${\bf g}_{1}={\bf g}_{1}^{+}\oplus {\bf g}_{1}^{-}$
falls in two parts corresponding to positive and negative odd roots
(with isospins $\pm\hf$ according to $su(2)_\theta$, cf. \cite{imp}).
Any basis for ${\bf g}_{1}^{+}$ is indexed by the positive odd roots
$\beta\in\Delta_+^1$. We denote the $su(2)^+\oplus su(2)^-$ transformation
of such a generic element $X_\beta$, by
\ben
t(X_{\beta})=(tX)_{\beta}=\sum_{\beta'\in\Delta^1_+}t_{\beta,\beta'}
X_{\theta-\beta'}
\een
We now turn to a brief description of the classical superconformal
algebra, $\At^{cl}$, to be compared below with the well known quantum $\At$,
\cite{gptvp,vp,gs,stvp88}:

Besides the energy-momentum tensor, $L(z)$, with (classical) central charge
$c_{cl}=-6k$,
it has the following generators:
Two sets of commuting affine $\widehat{SU}(2)^\pm$ currents, related to
the currents of the (classical)
even subalgebras of the affine Lie superalgebra
$\{J_{\alpha_i} \ , \  H^i\}$, with $\alpha_i\in\Delta (G)$,
for $i=2,3$ with (classical) level, $k$.
\footnote{Notice that $k$ is negative for a unitary representation
in our notation
since we are using negative definite metric for the even roots. We change to
a more standard notation presently.}

Denote
$H_{\alpha_i}\equiv\frac{2\alpha_i\cdot H}{\alpha_i^2}$
and write
$$\vec{J}_i\equiv (J_{\alpha_i},J_{-\alpha_i},H_{\alpha_i})
=(T^{\pm +},T^{\pm -},2T^{\pm 3})\equiv \vec{J}^\pm,$$
for $i=2,3$ . Introduce
$\vec{t_i}\cdot\vec{J}_i
\equiv\hf [  t^{\alpha_i}J_{-\alpha_i}+
            t^{-\alpha_i}J_{\alpha_i}+t^iH_{\alpha_i}]$
and
$
\vec{J}^2_i\equiv\hf [ J_{\alpha_i}J_{-\alpha_i}+ J_{-\alpha_i}J_{\alpha_i}
+H_{\alpha_i}^2],
$ (normal ordering of modes implied).
These affine currents satisfy
\bea
J_{\alpha_i}(z)J_{-\alpha_i}(w)&=&\frac{\frac{2k}{\alpha^2}}{(z-w)^2}
+\frac{H_{\alpha_i}(w)}{z-w}
+\ldots, \nn
H_{\alpha_i}(z)H_{\alpha_i}(w)&=&\frac{\frac{4k}{\alpha_i^2}}{(z-w)^2}
+\ldots, \quad
H_{\alpha_i}(z)J_{\pm\alpha_i}(w)=\pm\frac{2J_{\pm\alpha_i}(w)}{z-w}
+\ldots,
\eea
and they are conformal primaries with dimension 1.
All these OPE's are understood here in the classical case in the formal sense
of the hamiltonian structure.

Next there are four supercurrents, $G_\beta (z)$
indexed by the positive odd roots,
related to the notation $(+,-,+K,-K)$ of \cite{stvp88,gptvp} as above.
They are conformal primaries with dimension, $3/2$, and affine primaries
transforming as ($i=2,3$)
\ben
J_{\alpha_i}(z)G_{\beta}(w)=\frac{(t^{\alpha_i}G)_{\beta}(w)}{z-w}
+\ldots, \quad
H_{\alpha_i}(z)G_{\beta}(w)= \frac{(t^{i}G)_{\beta}(w)}{z-w}+\ldots .
\een
The general result of \cite{imp} for the classical OPE's of the supercurrents
reduces in this case to
\bea
G_{\theta -\beta'}(z)G_{\beta}(w)&=& -2k\delta_{\beta,\beta'}(z-w)^{-3}
+2(z-w)^{-2}M_{\theta-\beta',\beta}(w)
+(z-w)^{-1}\partial_wM_{\theta-\beta',\beta}(w)\nn
 &+&  \!\!\!\!
\delta_{\beta,\beta'}(z-w)^{-1}\left[ L(w)+
\frac{1}{K}\{\vec{J}^2_{2}+\vec{J}^2_{3}
+t^2_{\theta-\beta,\beta}H_{\alpha_2}(w)H_{\alpha_3}(w)\} \right]
+\ldots,
\eea
where
\ben
M_{\theta -\beta',\beta}\equiv \left [\gamma \vec{t}_2\cdot\vec{J}_2+
(1-\gamma)\vec{t}_3\cdot\vec{J}_3\right ]_{\theta -\beta',\beta}.
\een

It is convenient to reparametrize the two constants, $k,\gamma$ in terms of
$K^\pm$ (and $K\equiv K^++K^-$) thus
\ben
\gamma=\frac{K^-}{K}, \quad 1-\gamma\equiv\frac{K^+}{K}, \quad
k= -\frac{K^+K^-}{K},
\een
showing that the parameters, $K^{\pm}$, are simply the classical
levels of the two commuting, affine $\widehat{SU}(2)$'s.

The {\em quantum} algebra, $\At$, has an entirely similar form with the OPE's
having the standard meaning in conformal field theory. However, some of the
parameters get ``renormalized". Thus the energy momentum tensor has
central charge \cite{gptvp,imp}
\ben
c=-6k-3=6\frac{K^+K^-}{K}-3.
\een
The two commuting affine
$\widehat{SU}(2)$'s have levels $\tilde{K}^\pm$ rather than
$K^\pm$, where $\tilde{K}^\pm\equiv K^\pm-1$.
Similarly, the operator, $M_{\theta -\beta',\beta}$ is modified into
\ben
M_{\theta -\beta',\beta}=
\left [ \frac{\tilde{K}^-}{K}\vec{t}^+\cdot\vec{J}^++
\frac{\tilde{K}^+}{K}\vec{t}^-\cdot\vec{J}^-\right ]_{\theta -\beta',\beta}.
\een
We now turn to the explicit free field representation of this quantum
algebra, cf. \cite{imp}.
We introduce commuting sets of affine currents ($i=2,3$),
\ben
\vec{\hat{J}}_i\equiv (\hat{J}_{\alpha_i},\hat{J}_{-\alpha_i},
\hat{H}_{\alpha_i}),
\een
with similar OPE's as the currents $\vec{J}_i$,
except that they correspond to
Kac-Moody levels, $\tilde{K}^\pm-1$ rather than $\tilde{K}^\pm$. We could
introduce the standard Wakimoto free field representations \cite{wakimoto}
for those as well,
but we shall not need these explicitly. We further introduce free fermions,
$\chi_{\beta}(z)$, with OPE's
\ben
\chi_{\theta -\beta '}(z)\chi_{\beta}(w)=
\frac{\delta_{\beta '\beta}}{z-w}+\ldots .
\een
They are taken to commute with the currents, $\vec{\hat{J}}_i(z)$,
$i=2,3$. Finally
a free scalar, $\phi(z)$ coupled to world sheet curvature is used. Based on
these we first introduce fermionic affine
$\widehat{SU}(2)_1\times \widehat{SU}(2)_1$ currents
\ben
\vec{J}_{F,i}(z)\equiv \sum_{\beta,\beta'\in\Delta^1_+}
\chi_{\theta -\beta}(z)
(\vec{t}_{\beta ,\theta -\beta'})_i\chi_{\beta'}(z).
\een
Following ref. \cite{imp} we write (the dual Coxeter number, $h^\vee$ of
the exceptional Lie superalgebra, $D(2|1;\alpha)$, is zero)
\bea
\alpha_+&\equiv&\sqrt{k+h^\vee}=-i\sqrt{\frac{K^+K^-}{K}},\nn
Q&\equiv&\sqrt{\theta^2}(\frac{2-|\Delta^1_+|}{2\alpha_+}-
\frac{2\alpha_+}{\theta^2})=-\sqrt{2}(\alpha_+^{-1}+\alpha_+).
\eea
Also
\ben
L_\phi=-\hf(\partial \phi)^2-\frac{iQ}{2}\partial^2\phi,\ \
L_F = \hf\sum_{\beta\in\Delta^1_+}\partial\chi_{\theta -\beta}
\chi_{\beta},\ \
L_{\hat{J}^\pm}=\frac{2}{K^\pm}(\vec{\hat{J}}^\pm)^2 .
\een
It is then a straightforward (albeit tedious) task to check that the
following provides the sought free field realization of the algebra:
\bea
L(z)&=&L_{\hat{J}^+}(z)+L_{\hat{J}^-}(z)+L_F(z)+L_\phi (z), \nn
\vec{J}^\pm (z)&=&\vec{\hat{J}}^\pm (z)+\vec{J}_F^\pm (z),  \\
G_{\beta}(z)&=&\{\frac{Q}{\sqrt{2}}\partial+\frac{i}{\sqrt{2}}\partial\phi
+\frac{2\gamma}{\alpha_+}
(\vec{\hat{J}}^++\frac{3}{2}\vec{J}_F^+)\cdot\vec{t}^+
+  \frac{2(1-\gamma)}{\alpha_+}
(\vec{\hat{J}}^-+\frac{3}{2}\vec{J}_F^-)\cdot\vec{t}^-\}\chi_\beta(z).
\nonumber
\eea

It is easy to verify that these expressions generalize the ones previously
presented in the literature. Thus, for $\gamma=\hf$ or $K^+=K^-$, the
algebra $\At$ reduces to the $N=4$ superconformal algebra with an affine
$\widehat{SO}(4)$, and the above expressions
to the ones given by Mathieu \cite{mathieu}.
In particular one may easily verify that
$\vec{\hat{J}}_F^+\cdot\vec{t}^+\chi_\beta (z)
\equiv -\vec{\hat{J}}_F^-\cdot\vec{t}^-\chi_\beta (z)$
so that these (triple fermionic)
terms cancel in that limit.

Second, in the limit
$\gamma\rightarrow 1$, or $K^-\rightarrow\infty$, or
$\alpha_+\rightarrow -i\sqrt{K^+}$ (with a similar
``charge exchanged" limit), the $\At$ algebra reduces to the singly extended
$N=4$ algebra with a single affine $\widehat{SU}(2)$.
In that limit the above expressions
reduce to the ones recently given by Matsuda \cite{matsuda}.

These results have immediate consequences for the representation theory of
the $\At$ algebra, as studied  in \cite{gptvp,pt,opt}. In particular, it
follows that characters of highest weight representations (for
details of the definition of $\At$ characters, see \cite{pt}) take the form
\ben
Ch^{\At}(h,K^+,K^-,\lp ,\lm ;q,z_+,z_-)=\chi_\phi (q)\chi_F(q,z_+,z_-)
\chi_{2\lp}^{\tilde{K}^+-1}(q,z_+)\chi_{2\lm}^{\tilde{K}^--1}(q,z_-)
\een
where
\ben
\chi_\phi (q)=\frac{q^{h_\phi +\frac{Q^2}{8}}}{\eta (q)}
\een
is the character of the $\phi$-module with
$c_\phi = 1-3Q^2$
and conformal dimension, $h_\phi$.
\ben
\chi_F(q,\xi_+,\xi_-)=\frac{\theta_3(q,z_+)\theta_3(q,z_-)}{\eta^2(q)}
\een
is the contribution from the 4 fermions. For simplicity we only give
this expressions
in the NS sector, but the translation to the
R sector is trivial using standard
methods, cf. \cite{defev}. Finally
$\chi_{2\ell}^k(q,z)$
is the standard affine $\widehat{SU}(2)_k$
character for isospin $\ell$ and level $k$.
This expression for the character is in complete agreement with what
was found in ref. \cite{pt} for the characters of unitary, irreducible,
{\em massive} representations.
In \cite{pt} it looked miraculous (i) that the factorized form should
appear, and (ii) that affine $\widehat{SU}(2)$ characters at levels
$\tilde{K}^\pm-1$ should occur. These aspects now are fully explained.

As shown in \cite{gptvp} unitary representations of $\At$ occur only
when (a) we have integrable representations of affine
$\widehat{SU}(2)_{\tilde{K}^+}\times \widehat{SU}(2)_{\tilde{K}^-}$
(i.e.: $K^\pm$
are positive integers, $2\lp ,2\lm$ are positive integers
$\leq \tilde{K}^\pm$ in the NS sector),
and (b) when the conformal dimension in the NS sector is bounded by
\ben
h\geq h_0^{NS}\equiv\frac{1}{K}\{(\lp -\lm )^2+K^+\lm +K^-\lp\}.
\een
The corresponding bound in the R sector is given by
\ben
h\geq h_0^R\equiv\frac{1}{K}(\lp +\lm)^2+\frac{K^+K^-}{4K}-\frac{1}{4}.
\een
At the limiting value, the massive representation is no longer
irreducible, but splits into two massless or chiral representations.

Clearly, for $h_\phi =-\frac{Q^2}{8}$ (which is positive for integrable
representations) the characters transform according to a finite dimensional
representation of the modular group.

We now turn to the determination of screening operators relevant for the
discussion of null states in degenerate representations. Again, these will
turn out to be closely related to but generalizing the known expressions
\cite{mathieu,matsuda} for the two standard versions of $N=4$
superconformal algebras. Screening operators associated with the affine
$\widehat{SU}(2)_{\tilde{K}^+-1}\times \widehat{SU}(2)_{\tilde{K}^--1}$
are standard
\cite{wakimoto}, and will not
be further discussed here (cf. also \cite{imp}).
In addition to those we find two other screening
operators (as in \cite{imp}).

We first introduce the primary multiplet of fields,
$$\Phi^{\overline{\Lambda}^\ast}_{\overline{\lambda}}(z)
\equiv[\Phi^{\hf}_{m^+}(z)]^{(+)}[\Phi^{\hf}_{m^-}(z)]^{(-)}$$
of the currents $\vec{\hat{J}}_i$, carrying the
$\overline{\Lambda}^\ast=(\hf ,\hf)$ representation of
$SU(2)^+\times SU(2)^-$,
with weight $\overline{\lambda}=(m^+,m^-) =(\pm\hf ,\pm \hf )$. These
weights may naturally be identified with the positive, odd roots, and
phases may be chosen so that
\ben
t^{\alpha}(\Phi^{\overline{\Lambda}^\ast}_{\beta}(z))=
\Phi^{\overline{\Lambda}^\ast}_{\beta +\alpha}(z)
\een
for $\alpha\in\Delta (G)$ and $\beta ,\beta +\alpha\in\Delta^1_+$. Then
the fermionic screening operator may be given as
\ben
S_f(z)=\sum_{\beta\in\Delta^1_+}\chi_{\beta}(z)
\Phi^{\overline{\Lambda}^\ast}_{\theta -\beta}(z)
\exp (-\frac{i\sqrt{\theta^2}\phi(z)}{2\alpha_+}),
\een
in accord with the structure of the
general formula in \cite{imp}. A detailed calculation
shows that
\ben
G_{\beta}(z)S_f(w)=\partial_w\left (
\frac{\Phi^{\overline{\Lambda}^\ast}_{\beta}(w)
\exp (-\frac{i\sqrt{\theta^2}\phi(w)}{2\alpha_+}) }{z-w}\right )
+\ldots.
\een
In comparison with the result in ref.\cite{matsuda} it may be noticed
that the sum over $\beta$ involves 4 terms which pairwise form singlets
of either $SU(2)^+$ or $SU(2)^-$ but not of both.
In the limit  $K^-\rightarrow\infty$ (say)
where one of these affine algebras are frozen out, it is only important
to have singlets of $\widehat{SU}(2)_{\tilde{K}^+}^+$ so that one actually
finds two independent fermionic screening operators in that limit as in
ref. \cite{matsuda}. In general, however, there
is only the single one we have provided. Nevertheless the same number of
families of null states are generated by our single chiral screening
operator as with the two of \cite{matsuda}, cf. the discussion of the Kac
determinant below.

As argued in \cite{imp} we expect another screening operator of the
form
\ben
S_{\theta}(z)=s_{\theta}(z)
\exp (\frac{2i\alpha_+\phi(z)}{\sqrt{\theta^2}}),
\een
in order to characterize the singular vectors for the boson $\phi$.
The explicit form for $s_{\theta}(z)$ has been obtained previously only
in a limited number of cases. In the present one we find it may be
expressed (for example) as follows.
\bea
s_{\theta}(z)&=&(\alpha_+^2+1)\chi_{+}(z)\chi_{-}(z)
\chi_{+K}(z)\chi_{-K}(z)+(1-2\gamma)[L_\phi (z)
-\frac{i\alpha_+}{\sqrt{2}}\partial^2\phi(z)-L_F(z)]\nn
&-&\gamma (L_{\hat{J}^+}-2\vec{\hat{J}}^+\cdot\vec{J}_F^+)(z)+
(1-\gamma)(L_{\hat{J}^-}-2\vec{\hat{J}}^-\cdot\vec{J}_F^-)(z)
\eea
again this expression reduces to the corresponding ones in the limits,
$\gamma\rightarrow \hf , 1$, \cite{mathieu,matsuda},
and for the (``most non-trivial") OPE's one finds
\bea
G_{\beta}(z)S_{\theta}(w)&=&\partial_w\Bigl\{ (z-w)^{-1}\{ [
i\alpha_+ (2\gamma -1)(\partial +\frac{1}{\sqrt{2}}\partial \phi )
+2\gamma
(\vec{\hat{J}}^+\cdot\vec{t}^+-\frac{2}{3}(\alpha_+^2+1)
\vec{J}_F^+\cdot\vec{t}^+)\nn
&-&2(1-\gamma)(\vec{\hat{J}}^-\cdot\vec{t}^--\frac{2}{3}(\alpha_+^2+1)
\vec{J}_F^-\cdot\vec{t}^-)]\chi_{\beta}(w)\}
{\rm e}^{\frac{2i\alpha_+\phi(w)}{\sqrt{\theta^2}}} \Bigr\}
+\cdots.
\eea

The singular vectors produced by integrating the screening charges for the
affine
$\widehat{SU}(2)_{\tilde{K}^+-1}\times \widehat{SU}(2)_{\tilde{K}^--1}$
(which we did not explicitly provide) may rather easily be seen to
produce (i) for integrable representations, precisely the singular vectors
constructed and used in ref.\cite{pt} for building characters for the
unitary representations of the $\At$ algebra (generalizing the techniques
of \cite{et,kp}), and (ii) singular vectors associated with non-unitary,
non-integrable representations in complete analogy to the case of
the singly extended $N=4$ theory, cf. \cite{matsuda}. In addition to those,
there are singular vectors produced by the screening operators, $S_f(z)$ and
$S_{\theta}(z)$, just given. Again the situation is rather analogous to the
one discussed in the special case $\gamma\rightarrow 1$ by Matsuda
\cite{matsuda}.
In that case families of null states were conjectured based on computer
studies by Kent and Riggs \cite{kr} some time ago.
In \cite{matsuda} it was shown how these were exactly produced by the
corresponding screening operators.
Consistency of the characters demands that these extra singular
vectors lie entirely outside the unitarity domain\footnote{One of us (JLP)
is indebted to A. Kent for illuminating discussions.}.
Further, once the multiplicity
of the zeros in the Kac-determinant corresponding to all
these singular vectors are understood and found to saturate the bound
provided by the corresponding level, one may consider that the final part
of a rigorous proof has been provided for the characters, \cite{et,pt},
and the Kac determinant as well.

In the present letter we confine ourselves to presenting a natural proposal
for how the Kent and Riggs conjecture for the Kac determinant should
generalize for a general
value of the parameter, $\gamma\neq 1$. For simplicity we consider
only the R-sector as in \cite{matsuda}. It is straight forward to repeat
the treatment in the NS sector or in fact to use ``spectral flow" from the
R sector result given here, \cite{defev}. We attempt a notation making
the comparison with ref. \cite{kr} easy.

The structure of highest weight multiplets in the R sector is as follows
(for a complete discussion, cf. \cite{pt}):
The massive representation has 4 $SU(2)^+\times SU(2)^-$
multiplets with the same (lowest) value of the conformal dimension, $h$,
and isospins parametrized as
$$(\lp ,\lm -\hf), (\lp -\hf ,\lm), (\lp -1 ,\lm -\hf),
(\lp -\hf ,\lm -1).$$
The representation is referred to by means of the triple $(h,\lp ,\lm )$
in addition to the two parameters, $(k,\gamma)$, or equivalently,
$(K^+,K^-)$.
The 4 isospin multiplets are connected by the action of the 4 supercurrent
zero-modes. The two first members above are annihilated by
$(G_+)_0,(G_{+K})_0$ and $(G_+)_0,(G_{-K})_0$, respectively. The Kac
determinant is the determinant of the sesquilinear form of inner products of
states in the module built on the highest weight multiplet, and having
definite values of the quantum numbers, thus $h,\lp_3 ,\lm_3$ are increased
respectively by $n,\Delta \lp ,\Delta\lm$.
It is denoted by
$M_{n,\Delta\lp ,\Delta\lm}(h,k,\gamma,\lp,\lm)$.

Multiplicities of states in this
subspace are given by partition functions,
$P^4_\gamma (n,2\Delta\lp ,2\Delta\lm )$ which are simple to describe
provided all generators act freely. In that case they are generated as
follows
\bea
P^4_\gamma (q,z_+,z_-)&\equiv&\prod_{n=1}^\infty
\frac{(1+q^nz_+z_-)(1+q^nz_+z_-^{-1})(1+q^nz_+^{-1}z_-)
(1+q^nz_+^{-1}z_-^{-1})}
{(1-q^3)(1-q^nz_+^2)(1-q^nz_+^{-2})(1-q^nz_-^2)(1-q^nz_-^{-2})}\nn
&\times&\frac{(1+z_+^{-1}z_-)(1+z_+^{-1}z_-^{-1})}{(1-z_+^{-2})(1-z_-^{-2})}
\nn
&\equiv&\sum_{n,2\Delta\lp ,2\Delta\lm=1}^{\infty}
q^nz_+^{2\Delta\lp}z_-^{2\Delta\lm}P^4_\gamma (n,2\Delta\lp ,2\Delta\lm ).
\eea
In case the supercurrent with $SU(2)^+\times SU(2)^-$ weight
$(\hf\epsilon^+,\hf\epsilon^-)$
and level, $N$ does {\em not} act freely (i.e. is expressible in terms of
other operators), the multiplicity is provided by the modified functions
\bea
P^4_\gamma (q,z_+,z_-;N,\epsilon^+,\epsilon^-)&\equiv&
P^4_\gamma (q,z_+,z_-)(1+q^Nz_+^{\epsilon^+}z_-^{\epsilon^-})^{-1}\nn
&\equiv& \!\!\!\!\!\!\!
\sum_{n,2\Delta\lp ,2\Delta\lm =1}^{\infty}
q^nz_+^{2\Delta\lp}z_-^{2\Delta\lm}P^4_\gamma (n,2\Delta\lp ,2\Delta\lm
;N,\epsilon^+,\epsilon^-).
\eea
We further generalize the various
functions of \cite{kr} (corresponding to $\rho =\eta =0$ in their notation;
this is the R sector), the vanishing of which corresponds to the occurrence
of the singular states as described below. Thus we define
\bea
f_{p,q}(h,k,\gamma,\lp ,\lm )&=&k(k+1+4h)-(k p-q)^2+4
(\gamma(\lp)^2+(1-\gamma)(\lm)^2) , \nn
g_{p^\pm,q^\pm}(h,K^\pm,\ell^\pm)&=&K^\pm |p^\pm |
+2\mbox{sgn}(p^\pm)\ell^\pm-q^\pm ,\nn
h_{N,\epsilon^+,\epsilon^-}(h,k,\gamma ,\lp ,\lm)&=&4h-k(4N^2-1)+1+
8N(\gamma\epsilon^+\lp +(1-\gamma)\epsilon^-\lm )\nn
&-&\frac{4}{K}(\epsilon^+\lp -\epsilon^-\lm)^2 .
\eea
These are simply related to corresponding functions in \cite{kr} in the limit
$\gamma\rightarrow 1$. The precise form follows from the identification of
singular states presented below.

The corresponding conjecture for the Kac determinant takes the form
\bea
M_{n,\Delta\lp ,\Delta\lm}(h,k,\gamma,\lp ,\lm)&=&
A\prod_{p,q>0}[f_{p,q}(h,k,\gamma,\lp ,\lm)]^{P^4_\gamma(n-pq,2\Delta\lp,
2\Delta\lm)}\nn
&\times&\prod_{\stackrel{p^+=-\infty}{q^+=1}}^{\infty}
[g_{p^+,q^+}(h,K^+,\lp)]^{P^4_\gamma(n-|p^+|q^+,2\Delta\lp
+\mbox{sgn}(p^+)q^+,2\Delta\lm)}\nn
&\times& \prod_{\stackrel{p^-=-\infty}{q^-=1}}^{\infty}
[g_{p^-,q^-}(h,K^-,\lm)]^{P^4_\gamma(n-|p^-|q^-,
2\Delta\lp ,2\Delta\lm+\mbox{sgn}(p^-)q^-)}\nn
&\times&\prod_{\stackrel{N=1}{\epsilon^\pm=\pm 1}}^{\infty}
[h_{N,\epsilon^+,\epsilon^-}(h,k,\gamma,\lp ,\lm)]^{P^4_\gamma(n-N,
2\Delta\lp -\epsilon^+,2\Delta\lm -\epsilon^-;N,\epsilon^+,\epsilon^-)}\nn
&\times&\prod_{\epsilon^-=\pm 1}
[h_{0,1,\epsilon^-}(h,k,\gamma,\lp ,\lm)]
^{P^4_\gamma(n,2\Delta\lp -1,2\Delta\lm -\epsilon^-;1,\epsilon^+)}
\eea
notice that
\bea
h_{0,1,-1}(h,k,\gamma,\lp ,\lm)
&=&4(h-\{\frac{1}{K}(\lp +\lm)^2+\frac{K^+K^-}{4K}-\frac{1}{4}\})\nn
&=&4(h-h_0^R)
\eea
where (as above) $h_0^R$
is the massless bound on the conformal dimension in the R sector.
The factor $A$ is a normalization factor.

We now briefly indicate how the screening operators we have found give
rise to singular vectors which in turn give rise to the above factors in
the Kac determinant.

In our free field representation a generic state is a tensor product of
states pertaining to (i) the Fock space of the 4 free fermions, (ii) a
representation space of the two commuting affine currents,
$\vec{\hat{J}}^\pm$, and (iii) the scalar, $\phi$:
$ |F\rangle\otimes |\hat{\lp}\rangle\otimes |\hat{\lm}\rangle
\otimes | p\rangle$.
We refer to $p$ as the ``momentum", the corresponding state being created
by a vertex operator
$e^{i\sqrt{\theta^2}p\phi(z)}$.

We first consider singular vectors produced by $p$-fold integrals of
the  $\phi$-screening operator, $S_\theta$, carrying momentum, $\alpha_+$.
Consider acting with the $p$-fold integral on the state
$|0\rangle_R\otimes|\hat{\ell}^\pm=\ell^\pm-\hf\rangle\otimes
|P-p\alpha_+\rangle$.
Here the combined Ramond vacuum states $|0\rangle_R$ of the 4 fermions
provide
the multiplet structure in the R sector described above. They transform as
a $(\hf,\hf)$ representation and provides a contribution of $4/16$ to the
conformal dimension. The action of the $p$-fold integral of $S_\theta$
will produce a singular vector which itself is a descendant of a highest
weight vector with momentum, $P$, and isospins unchanged. The on-shell
condition that the level (increase in conformal dimension) is a positive
integer, $N=p\cdot q$ gives rise to the quantization of momenta
\ben
P=P_{p,q}=-\frac{1}{2\alpha_+}(q-1)+\frac{\alpha_+}{2}(p+1),
\een
which is a generalization of \cite{matsuda} and conforms with \cite{imp}
(after $P\rightarrow -Q-P$). The corresponding conformal dimension takes a
value given by the vanishing of
$f_{p,q}(h,k,\gamma,\lp ,\lm )$
and justifies the first factor in the Kac determinant. One also checks that
the corresponding values of $h$ lie below the unitarity bound, so that
these singular vectors do not affect the calculation of characters of
unitary representations, \cite{pt}.

The identification of singular vectors arising from the well known
screening charges (not repeated here) of the affine
$\widehat{SU}(2)_{\tilde{K}^+-1}\times \widehat{SU}(2)_{\tilde{K}^--1}$
is entirely analogous to
the singly extended case, \cite{matsuda}, and gives rise
to the second and third factor in the Kac determinant.
For integrable representations these are exactly the ones employed in the
construction of characters in ref. \cite{pt}.

Singular vectors produced by the fermionic screening operator, $S_f$,
are obtained by integrating that once only, and acting on a suitable state
(cf. \cite{katomatsuda}), which we take to be
$|0\rangle_R\otimes |\hat{j}^+,\hat{j}^-\rangle\otimes
|P+\frac{1}{2\alpha_+}\rangle$,
since the chiral screening operator has momentum, $-1/2\alpha_+$.
The result
will be a singular vector with $SU(2)^+\times SU(2)^-$ weight,
$(\lp_S,\lm_S)=(\hat{j}^++\hf,\hat{j}^-+\hf )$
(the extra $(\hf,\hf)$ coming as before from the Ramond ground state
multiplet
of the four fermions). The singular vector is a descendant of a highest
weight
state with momentum, $P$ and $SU(2)^+\times SU(2)^-$ weight
$(\lp ,\lm )=(\hat{j}^++\hf+\epsilon^+\hf,\hat{j}^-+\hf+\epsilon^-\hf)$
provided the on-shell condition that the level is a positive integer
(or zero) is fulfilled.
Here there are four possibilities corresponding to $\epsilon^\pm =\pm 1$.
The reason
is that the screening operator, $S_f$ involves the operators,
$[\Phi^{\hf}_{m^\pm}]^\pm$ transforming as $(\hf,\hf)$ primaries of the
currents, $\vec{\hat{J}}^\pm$. More explicitly the 4 possibilities may be
seen
by using standard free field representations for these operators. These
representations involve a vertex operator or it's dual and they give rise
to the
total of 4 possibilities. If the level is denoted by $N$ the quantization
condition for the momentum is
\ben
P=P_N=\alpha_+N+\hf \alpha_++\alpha_+\left [ \frac{\epsilon^+\lp -\hf}{K^+}
+\frac{\epsilon^+\lp -\hf}{K^+} \right ] .
\een
This gives rise to a value for the conformal dimension provided by the
vanishing  of the function
$h_{N,\epsilon^+,\epsilon^-}(h,k,\gamma ,\lp ,\lm )$
above. These conformal dimensions all lie below the unitarity bound, $h_0^R$,
except for $N=0, \epsilon^+=-\epsilon^-$ which is precisely the condition for
masslessness. Thus the existence of these states is also consistent with
the character calculation, \cite{pt}. Finally they give rise to the
remaining two factors in the Kac determinant.

The multiplicities here are given under the assumption that a certain
supercurrent  mode is redundant. Such phenomena are well known (see for
example \cite{et} in the singly extended $N=4$ case, and \cite{pt} in the
doubly extended $N=4$ case) in similar contexts, but in this letter we have
not in fact attempted a full understanding.\\[2cm]
{\bf Acknowledgements:}

One of us (JLP) is grateful to the groups at University of Durham,
Dept. of Mathematics, and at University of Cambridge, DAMTP for discussions
relevant to this work, in particular very useful discussions with
A. Kent and A. Taormina are gratefully acknowledged, as is partial
support under EEC grant SC1 0394-C(EDB).

\end{document}